\documentclass[showpacs,prl,twocolumn,aps,superscriptaddress,preprintnumbers]{revtex4}

\usepackage{epsfig}
\usepackage{graphicx}
\usepackage{amsmath}
\usepackage{amsfonts}
\usepackage{epstopdf}

\def\slashchar#1{\setbox0=\hbox{$#1$}     		
   \dimen0=\wd0                                 	
   \setbox1=\hbox{/} \dimen1=\wd1               	
   \ifdim\dimen0>\dimen1                        	
      \rlap{\hbox to \dimen0{\hfil/\hfil}}      	
      #1                                        	
   \else                                        	
      \rlap{\hbox to \dimen1{\hfil$#1$\hfil}}   	
      /                                         	
   \fi}

\renewcommand{\vec}{\boldsymbol}
\newcommand{\be}{\begin{equation}}
\newcommand{\ee}{\end{equation}}
\newcommand{\bear}{\begin{eqnarray}}
\newcommand{\eear}{\end{eqnarray}}
\newcommand{\ba}{\begin{array}}
\newcommand{\ea}{\end{array}}

\begin{document}

\title{Conformal anomaly as a source of soft photons in heavy ion collisions}

\author{G\"ok\c ce Ba\c sar}
\email{basar@tonic.physics.sunysb.edu}
\affiliation{Department of Physics and Astronomy, Stony Brook University, Stony Brook, New York 11794-3800, USA}

\author{Dmitri E. Kharzeev}
\email{Dmitri.Kharzeev@stonybrook.edu}
\affiliation{Department of Physics and Astronomy, Stony Brook University, Stony Brook, New York 11794-3800, USA}
\affiliation{Department of Physics,
Brookhaven National Laboratory, Upton, New York 11973-5000, USA}

\author{Vladimir Skokov}
\email{VSkokov@bnl.gov}

\affiliation{Department of Physics,
Brookhaven National Laboratory, Upton, New York 11973-5000, USA}

\date{\today}

\begin{abstract}
We introduce a novel photon production mechanism stemming from the conformal anomaly 
of QCD$\times$QED and the existence of strong (electro)magnetic fields in heavy ion collisions. 
Using the hydrodynamical description of the
bulk modes of QCD plasma, we show that this mechanism leads to the photon production yield that is comparable to the yield from
conventional sources. This mechanism also provides a significant positive contribution to the azimuthal anisotropy of photons, $v_2$,
as well as to the radial ``flow". We compare our results to the data from the PHENIX Collaboration. 
\end{abstract}

\maketitle


The manifestations of quantum anomalies in the collective dynamics of relativistic plasmas have attracted considerable interest recently. In the presence of background fields, triangle anomalies can lead to non-conservation of currents. In the case of the axial $AVV$ anomaly involving an axial $A_\mu$ and two vector $V_\mu$ currents, 
the presence of an external magnetic field and a finite density of chiral charge leads to the generation of electric current in QCD plasma  -- the chiral magnetic effect~\cite{Kharzeev:2004ey,Kharzeev:2007tn,Kharzeev:2007jp,Fukushima:2008xe,Kharzeev:2009fn}.  
At finite density of (vector) baryon charge, magnetic field induces the flow of axial current; this is the chiral
separation effect~\cite{son:2004tq,Metlitski:2005pr,son:2009tf}. Both of these phenomena are an integral part of relativistic hydrodynamics, and in fact are required by the second law of thermodynamics \cite{son:2009tf,Kharzeev:2011ds,Loganayagam:2011mu,Chapman:2012my}. 

 In this letter we investigate the related effects stemming from the {\it conformal} $SVV$ anomaly \cite{sa} that involves a scale (dilatational) current $S_\mu$ and two vector currents $V_\mu$ and reflects the violation of conformal invariance of QCD by quantum effects. The conformal anomaly results from the running of the coupling constant (asymptotic freedom \cite{GW}) 
 in QCD and expresses the non-conservation of the dilatational current $S^\mu$,  so that the trace of the energy-momentum tensor $\theta^\mu_\mu$ does not vanish even in the chiral limit of massless quarks: $\partial^\mu S_\mu = \theta^\mu_\mu$.
 The quarks carry both color and electric charges, so 
 when QCD is coupled to electromagnetism, the quark triangle diagram induces an anomalous coupling of the trace of the energy-momentum tensor to photons \cite{Ellis:1984jv,Crewther:1972kn,Chanowitz:1972vd}. The trace of the energy-momentum tensor in hydrodynamics excites the bulk modes of the fluid that are abundant in (non-conformal) quark-gluon plasma \cite{Kharzeev:2007wb,Karsch:2007jc,Meyer:2007dy,Romatschke:2009ng}. The heavy ion collisions at early times produce very strong background magnetic fields \cite{Kharzeev:2007jp,Skokov:2009qp}. As a result, the conformal anomaly acts as a source of photon production that is powered by the energy of the bulk hydrodynamical modes in the plasma. This is the mechanism of photon production that will be discussed in detail below. Note that while we will use hydrodynamics to describe the bulk modes in the plasma, the deviation from equilibrium in general need not be small for our mechanism to operate. For example, the non-equilibrated Bose-Einstein condensate of gluons 
 \cite{Blaizot:2011xf} may be even more effective in producing photons. Note that unlike in the conventional scenario, the quarks in our case appear only in the triangle loop that receives contributions from the virtual UV modes -- so the production of real on-shell quarks is not required, and the mechanism can operate even at very early times.

Because of the relatively weak interactions with the medium, ``direct" (i.e. not resulting from the hadron decays) photons play an important role of a ``thermometer" of the quark-gluon plasma \cite{Feinberg:1976ua,Shuryak:1978ij} since the rate of their production per unit volume is expected to scale with the temperature $T$ as $\sim T^4$. Recent measurements of the PHENIX collaboration 
show very large excess of direct photons with the transverse momentum up to 3 GeV in AuAu collisions at RHIC \cite{Adare:2011zr}. 
The azimuthal anisotropy (often called ``elliptic flow'') 
of the produced photons has also been measured and reported \cite{Adare:2011zr}. 
It has been found that the anisotropy is large and similar to the elliptic flow of hadrons. This result contradicts the current theory of photon production. Indeed, it has been expected (see e.g. \cite{Chatterjee:2007xk}) that the elliptic flow of direct photons would be much smaller than that of hadrons because a significant fraction of photons has to be produced at early times, when the temperature is the highest. The measured yield of soft photons is indeed large and requires an early time production mechanism (see \cite{vanHees:2011vb,Dusling:2009ej} and references therein). At these early times the hydrodynamical flow has not been built up yet, and so the photons produced at that time are not expected to possess a significant azimuthal asymmetry \cite{vanHees:2011vb}. Moreover, the jet-medium interactions and the resulting induced bremsstrahlung of photons is expected to lead to a negative contribution to the elliptic flow Fourier coefficient $v_2$ \cite{Turbide:2005bz} -- indeed, due to the geometry of the collision the produced medium has an almond shape and is elongated along the axis orthogonal to the reaction plane. 

Therefore the large and positive value of $v_2$ presents a serious challenge to theory.
Here we will consider the photon production mechanism stemming from the conformal anomaly of QCD$\times$QED and the 
presence of a high magnetic field in heavy ion collisions. We will demonstrate that this mechanism 
results in a significant photon and dilepton yields that are comparable to the ones from the ``conventional" mechanism and may potentially explain the $v_2$ puzzle for soft direct photons. 

Let us begin by reminding the basics of conformal anomaly.
In field theory the divergence of the dilatational current $S_\mu$ is equal to the trace of the energy-momentum tensor. In QCD, this divergence does not vanish signaling the breaking of scale invariance due to dimensional transmutation and the running coupling:
\be
 \partial^\mu S_\mu = \theta^\mu_\mu = \frac{\beta(g)}{2 g} G^{\mu\nu a} G_{\mu\nu a} + \sum_q m_q \left[ 1 + \gamma_m(g) \right] \bar{q} q,
 \ee
 where $\beta(g)$ is the beta-function of QCD, $m_q$ are the quark masses, and $\gamma_m(g)$ are the corresponding anomalous dimensions. The current $S_\mu$ acting on the vacuum produces scalar color-singlet states $\sigma$ of mass $m_\sigma$ with an amplitude $f_\sigma$:
 \be
 \langle 0 | S^\mu | \sigma \rangle = i q^\mu f_\sigma; \ \ \langle 0 | \partial_\mu S^\mu | \sigma \rangle =  m_\sigma^2\ f_\sigma.
 \ee

  \begin{figure}
 \includegraphics[scale=0.7]{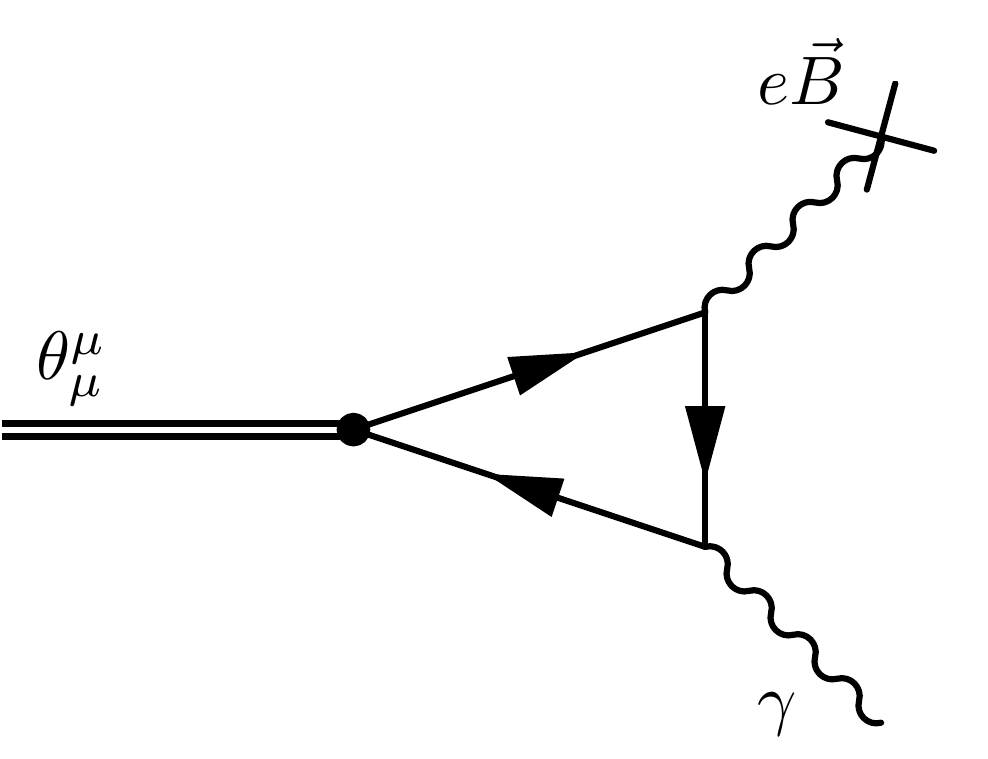}
 \caption{
 The coupling of the conformal anomaly to the external magnetic field resulting in 
 photon production. }
 \label{diagram}
 \end{figure}
	
	\vskip0.2cm
  Let us now consider the coupling of QCD scale anomaly to electromagnetism. This coupling can lead to the production of photons 
  in external magnetic field as described by the diagram of Fig.~\ref{diagram}.
   To evaluate the contribution of this diagram,
	we need to consider the coupling of the scalar meson to photons. 
  This coupling is described by the triangle quark diagram, and leads to the following effective interaction~\cite{Ellis:1984jv,Crewther:1972kn,Chanowitz:1972vd}: 
  \be
  {\cal L}_{\sigma \gamma \gamma} = g_{\sigma \gamma \gamma}\ \sigma\ F_{\mu\nu} F^{\mu\nu},
  \ee
  where $g_{\sigma \gamma \gamma}$ is related to the decay constant $f_\sigma$ discussed above and to the ratio of cross sections of $e^+e^-$ annihilation into hadrons and muons
  \be
  R \equiv \frac{\sigma(e^+e^- \to \gamma^* \to hadrons)}{\sigma(e^+e^- \to \gamma^* \to \mu^+\mu^-)}
  \ee 
  by 
  \be
  f_\sigma g_{\sigma \gamma \gamma} = \alpha\ \frac{R}{6 \pi},
  \ee
  where $\alpha$ is the fine structure constant. 
  The resulting width of $\sigma$ decay into two photons is given by \cite{Ellis:1984jv,Crewther:1972kn,Chanowitz:1972vd}
  \be\label{decay}
  \Gamma(\sigma \to \gamma \gamma) = g_{\sigma \gamma \gamma}^2 \ \frac{m_\sigma^3}{4 \pi} = \left(\frac{\alpha R}{3 \pi f_\sigma}\right)^2\ \frac{m_\sigma^3}{16 \pi}.
  \ee
  Using $R=5$ for six quark flavors (all of which contribute to the triangle diagram) and the values $m_\sigma = 550$ MeV, $f_\sigma = 100$ MeV discussed above, we get from 
  (\ref{decay}) the value $\Gamma(\sigma \to \gamma \gamma) \simeq 5$ KeV. This is in the middle of the range ($2 \div 10$ KeV) for the two photon decay width of $f_0(600)$ meson listed by PDG \cite{Nakamura:2010zzi}, supporting the identification of the lightest $\sigma$ dilaton with this meson. This allows us to fix the value $g_{\sigma \gamma \gamma} \simeq 0.02\ {\rm GeV}^{-1}$. Now we have all the information necessary to evaluate the diagram of Fig.~\ref{diagram}.
    \vskip0.2cm

	To compute the photon production rate from the diagram of Fig.~\ref{diagram}, we evaluate the imaginary part for the photon self-energy, see ~\cite{Weldon:1983jn,McLerran:1984ay}.
	A straightforward calculation yields for the production rate at mid-rapidity ($q_z=0$) the following expression:
	\begin{eqnarray}
	q_0 \frac{d \Gamma_B}{d^3 q} &=& 2 \left(  \frac{g_{\sigma \gamma \gamma} } { 
	\pi f_\sigma m_\sigma^2   } \right)^2 \times
	\nonumber 
	\\ &&
	\frac{ (B_y^2- B_x^2) q_x^2+q_\perp^2 B_x^2}{\exp(\beta q_0)-1} \rho_\theta  (q_0 =  |\vec{q}| ).
	\label{Anomalous}
	\end{eqnarray}
Since we consider production of photons in the QCD plasma, it is appropriate to use the hydrodynamic spectral function of the bulk mode $\theta$ \cite{Hong:2010at,Meyer:2010ii}:
\begin{eqnarray}
\rho_\theta(q_0,\vec q)=\frac{1}{\pi}\mathcal{I}m[G_R^{\mu\mu,\nu\nu}\,(q_0,\vec q)]=9q_0\frac{\zeta}{\pi} +\nonumber\\
\frac{9}{\pi} (\epsilon+p)\,\left(\frac{1}{3}-c_s^2\right)^2\,\frac{q_0\Gamma_s\,\vec q^4}{(q_0^2-c_s^2\,\vec q^2)^2+(q_0\Gamma_s \vec q^2)^2},
\end{eqnarray}
where $\Gamma_s=(4/3\eta+\zeta)/(\epsilon+ p)$ is the sound attenuation length and $\eta$ and $\zeta$ are shear and bulk viscosities. The second term describes the sound peak at $q_0=c_s |\vec q|$. 
The sound mode does not contribute to the production of real photons since the width of the sound peak is not large enough to 
reach the null dispersion of photons. 
Therefore  the photon production is dominated by the bulk viscosity $\zeta$:
\begin{equation}
	\rho_\theta (q_0 =  |\vec{q}| )\approx \frac{9 q_0} {\pi}  \zeta, 
	\label{rho_approx}
	\end{equation}

	In deriving Eq.~(\ref{Anomalous}) we neglected the $z$-component of the magnetic field, because it is expected 
	to be an order of magnitude smaller than $B_x$ and $B_y$ ($B_z\sim B_{x,y}/\gamma$); we also neglect the 
	contribution of the electric field.
	
	In what follows we will compare our result with the baseline provided by the conventional {
	thermal} photon production
	rate {
	recently calculated on lattice}~\cite{Ding:2010ga}:
	\begin{equation}
	q_0 \frac{d \Gamma}{d^3 q} = \frac{C_{\rm em} \alpha_{\rm em}}{4 \pi^2} \frac{\rho_V (q_0 =  |\vec{q}| )} {\exp(\beta q_0)-1} ,
	\label{Conventional}
	\end{equation}
  where  $C_{em} = \frac{e^2}{3} R \equiv \sum_f Q_f^2$ with $Q_f$'s are the electric charges of the quarks, and 
	$\rho_V$ is the vector current spectral function that in the limit of $q_0\to0$ and $\vec{q}\to0$ is
	related to the electric conductivity:
	\begin{equation}
		\sigma_{\rm em} = \frac{C_{\rm em}} {6} \lim\limits_{q_0\to0} \frac{\rho_V ( q_0 ,  |\vec{q}|=0  )}{q_0}.
		\label{conductivity}
	\end{equation}
{
Note that this conventional mechanism (\ref{Conventional}) is expected to be the dominant one for low transverse momentum, $p_\perp$, photons. For photons with $p_\perp\sim$ 2 GeV and above there will be additional contributions to the rate which can be calculated perturbatively. However we did not include these additional contributions as we are mainly interested in low $p_\perp$ photons.}

	The spectral function for $\theta$ and the bulk viscosity was calculated in lattice QCD ~\cite{Meyer:2007dy,Meyer:2010ii}. However the extraction of bulk viscosity from the lattice data is notoriously difficult. 
  To get an independent estimate of the bulk viscosity we thus follow~\cite{Weinberg:1971mx,Arnold:2006fz} and assume that 
	\begin{equation}
	\frac{\zeta}{\eta} = C_\zeta \left(\frac{1}{3} - c_s^2 \right)^2. 
	\label{bulk}
	\end{equation}
	{
	Thus the bulk viscosity vanishes in the conformal limit, $c_s^2=1/3$.  
	In the relaxation time approximation, 
	this expression is obtained in the kinetic theory with $C_\zeta=15$ (see e.g.~\cite{Dusling:2011fd}).}  
	The paper~\cite{Dusling:2011fd} contains also a phenomenological estimate of the value 
of bulk viscosity inferred from the comparison of viscous hydrodynamical computations with the data on the elliptic 
flow of mesons and baryons. The resulting estimate is $\zeta/s = 0.005$~\cite{Dusling:2011fd}. 
Using the lattice data for the speed of sound 
in the freeze-out temperature range from Ref.~\cite{Borsanyi:2010cj}, $c_s^2 = 0.175 \div 0.221$, we infer for the bulk viscosity from (\ref{bulk}) the value 
of $C_\zeta = 2.5 \div 5$. The leading log calculations in SU(3) Yang Mills theory results in a much larger value $C_\zeta \simeq 48$, see Ref.~\cite{Dusling:2011fd}.
	In our calculations, we choose the 
 lowest value available in the literature, $C_\zeta = 2.5  \div 5$, with an assumption $\eta/s=1/4\pi$.

 \begin{figure}[t]
 \includegraphics[scale=0.3]{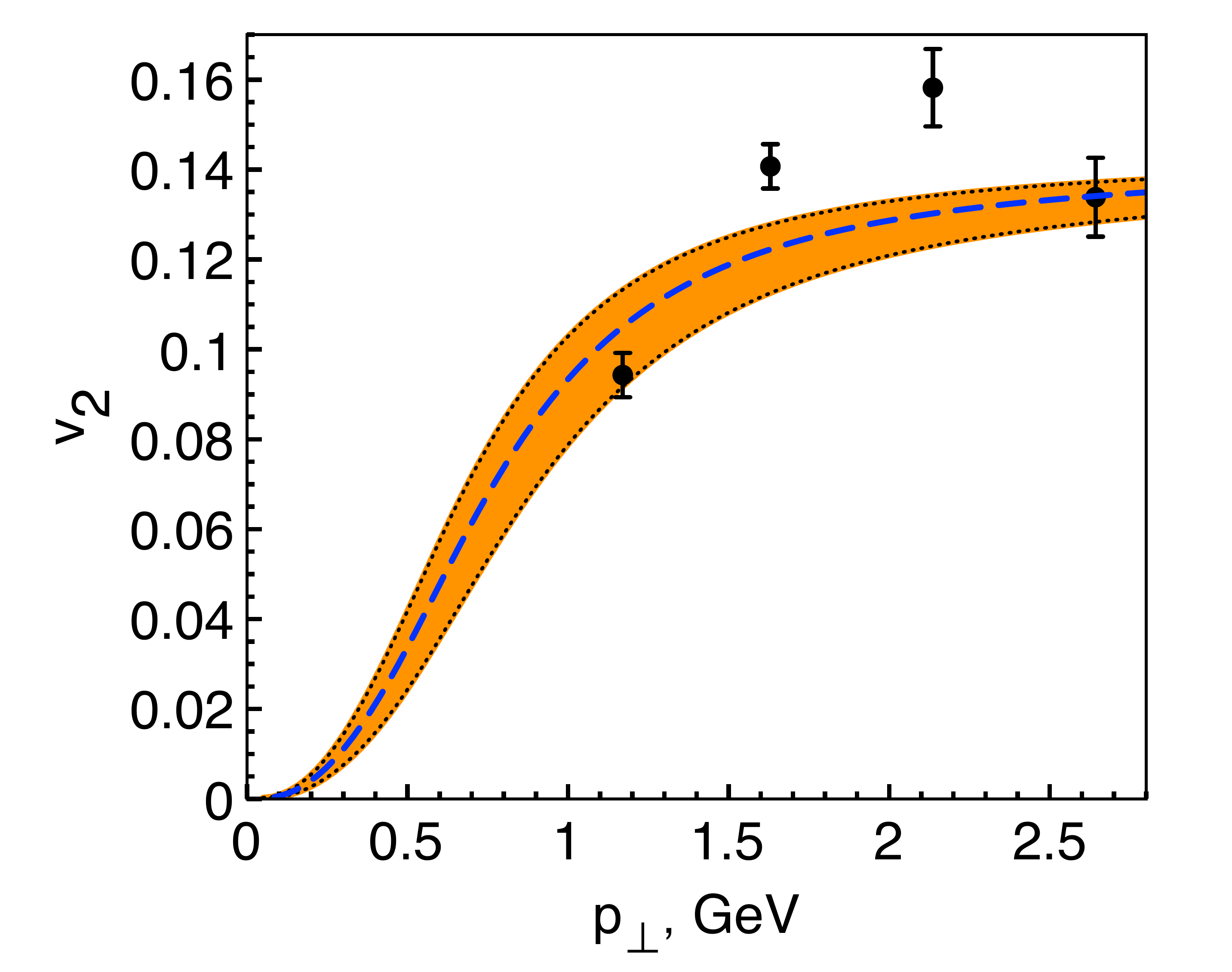}
 \caption{
 The azimuthal anisotropy $v_2$ of the direct photons for different values of bulk viscosity corresponding to $C_\zeta$ in the range of $2.5\div5$ calculated for minimum bias Au-Au collisions. The dashed line represents the results with $C_\zeta=4$. 
 The black dotes are the data from the PHENIX collaboration~\cite{Adare:2011zr} for minimum bias Au-Au collisions 
 at $\sqrt{s}=200$ GeV.
 }
 \label{band}
 \end{figure}

\vskip0.2cm
	The magnetic field in heavy ion collisions was estimated in Refs.~\cite{Kharzeev:2007jp} and \cite{Skokov:2009qp}; 
	the fluctuations of magnetic field were evaluated  in Refs.~\cite{Bzdak:2011yy} and \cite{Deng:2012pc}.
  In this paper, we neglect the spatial gradients of magnetic field and estimate the time dependence in the eikonal approximation taking into account only the (leading at large times) contribution from spectators:
	\begin{equation}
	 e B_{x,y}  ( t ) \simeq \frac{e B^0_{x,y}}{1+ (t/t_B)^2},  
	\label{eB}
	\end{equation}
  where $eB^0_i$ it the magnitude of the $i$-th component of the magnetic field 
	at $t=0$ and $t_B$ is the characteristic decay time. 
	The $x$-component of magnetic field at $t=0$, $B^0_{x}$,  is approximately 
	independent of the impact parameter $b$, while the $y$-component is linear in $b$. Both 
	components $B^0_{x,y}$ are linear as  a function of the collision energy,  $\sqrt{s}$; 
	the typical decay time is inversely proportional to $\sqrt{s}$.

  Here we neglect the transverse expansion of the fireball and assume that it has  
	an almond shape with the following characteristic sizes in $x$ and $y$ direction:  
	$l_x=(R_A-b/2)$ and $l_y=\sqrt{R_A^2-b^2/4}$, where $R_A$ is the radius of the colliding nuclei.   
	We approximate the time evolution of the temperature at early times using the Bjorken hydrodynamics 
	$T/T_0 = (\tau_0 / \tau )^{1/3}$, where $T_0$ is the initial temperature and $\tau_0$ 
	is the initial  time (given by the characteristic thermalization time of the gluons) that can be estimated 
	 in terms of the saturation scale, $Q_s$, and  the coupling constant, $\alpha_s$, see e.g. 
	Ref.~\cite{Blaizot:2011xf}.
	For Au-Au collisions at $\sqrt{s}=200$ GeV  we use $\tau_0=0.1$ fm/$c$.
	
To evaluate the bulk viscosity~(\ref{bulk}) we need the speed of sound, $c_s$ and the entropy, $s$; we use
	the model parametrization~\cite{Dumitru:2010mj}
	of lattice results for pure glue SU(3) theory.   Note that the transport coefficients of the plasma may be affected by magnetic field; for
	recent examples, see \cite{Tuchin:2011jw} and \cite{Basar:2012gh}.

\begin{figure}[t]
 \includegraphics[scale=0.3]{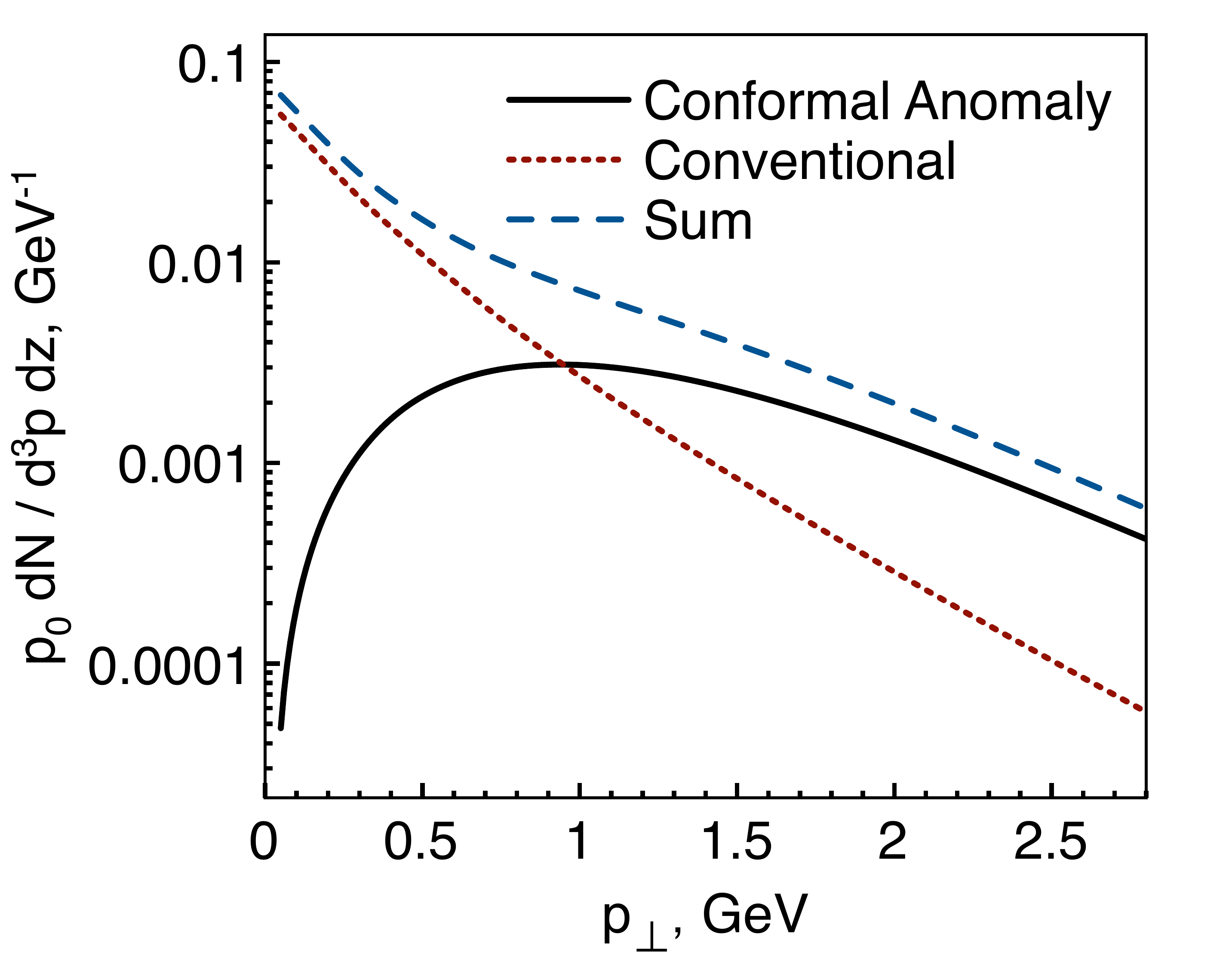}
 \caption{
 The transverse momentum spectra of the produced direct photons for $C_\zeta=2.5$ calculated for minimum bias Au-Au collsisions, see text for details.
 }
 \label{pt}
 \end{figure}

	Our results for the azimuthal anisotropy of photons calculated using both conventional production 
	mechanism and the one from the 
	conformal anomaly are shown in Fig.~\ref{band} for the minimum bias Au-Au collisions at $\sqrt{s}=200$~GeV. In our approximation (no transverse flow), the conventional mechanism does not give any 
	contribution to the azimuthal anisotropy. The comparison with the 
	experimental data from PHENIX~\cite{Adare:2011zr} indicates that conformal anomaly could account for a large fraction of
	the observed photon anisotropy.         
	
	In Fig.~\ref{pt} we show our result for the transverse momentum spectrum of direct photons. Due to the factor of $q^2$ in the production 
	rate~(\ref{Anomalous}), the spectrum of photons produced 
	due to conformal anomaly is enhanced in comparison to the conventional one
	at transverse momenta $k_\perp > 1$ GeV. The factor of $q^2$ in the rate hardens the transverse momentum spectrum, and magnetic field 
	grows with the impact parameter of the collision; these two effects thus conspire in mimicking both the elliptic and radial flow  
	 of photons in non-central collisions.   

An interesting corollary of our mechanism is the polarization of the produced photons (and low-mass dilepton pairs) relative to the  reaction plane 
of the collision. Other tests include the study of U-U collisions, where the deformed shape of the U nucleus may allow to separate \cite{Voloshin:2010ut} the eccentricity of the initial condition from the magnitude of magnetic field that drives our effect.

The calculations performed in the current paper are quite schematic and rely on a number of crude approximations. 
The bulk viscosity and its temperature dependence in SU(3) Yang-Mills theory are the major sources of uncertainty 
in our calculations. {
Nevertheless, in this letter we preferred to err on the side of caution and used the most conservative estimates 
for the bulk viscosity and other input parameters. In spite of this, we find that the quantum anomaly is responsible for a very substantial contribution to 
the overall soft photon yield.}
Realistic calculations treating the 3D hydrodynamical expansion and proper initial conditions
are required to reach a definite conclusion, and to compare to the data on the transverse momentum 
spectra of photons. These calculations are in progress and will be presented elsewhere.


\vskip0.2cm

We thank  A.~Bzdak, A.~Drees, J.~Ellis, B.~Jacak, J.~Liao, L.~McLerran, R.~Pisarski and H.-U.~Yee for discussions. D.~K. is grateful to CERN Theory Division for 
hospitality during the completion of this work. 
This research was supported by the US Department of Energy under Contracts DE-AC02-98CH10886 and DE-FG-88ER41723.



\end{document}